\def \SAIT #1 #2 {{\em Mem.\ Soc.\ Astron.\ It.\/} {\bf #1}, #2}
\def \MESS #1 #2 {{\em The Messenger\/} {\bf #1}, #2}
\def \ASTRNACH #1 #2 {{\em Astron. Nach.\/} {\bf #1}, #2}
\def \AAP #1 #2 {{\em Astron. Astrophys.\/} {\bf #1}, #2}
\def \AAL #1 #2 {{\em Astron. Astrophys. Lett.\/} {\bf #1}, L#2}
\def \AAR #1 #2 {{\em Astron. Astrophys. Rev.\/} {\bf #1}, #2}
\def \AAS #1 #2 {{\em Astron. Astrophys. Suppl. Ser.\/} {\bf #1}, #2}
\def \AJ #1 #2 {{\em Astron. J.\/} {\bf #1}, #2}
\def \ANNREV #1 #2 {{\em Ann. Rev. Astron. Astrophys.\/} {\bf #1}, #2}
\def \APJ #1 #2 {{\em Astrophys. J.\/} {\bf #1}, #2}
\def \APJL #1 #2 {{\em Astrophys. J. Lett.\/} {\bf #1}, L#2}
\def \APJS #1 #2 {{\em Astrophys. J. Suppl.\/} {\bf #1}, #2}
\def \APSS #1 #2 {{\em Astrophys. Space Sci.\/} {\bf #1}, #2}
\def \ASR #1 #2 {{\em Adv. Space Res.\/} {\bf #1}, #2}
\def \BAIC #1 #2 {{\em Bull. Astron. Inst. Czechosl.\/} {\bf #1}, #2}
\def \JSQRT #1 #2 {{\em J. Quant. Spectrosc. Radiat. Transfer\/} {\bf #1}, #2}
\def \MN #1 #2 {{\em Mon. Not. R. Astr. Soc.\/} {\bf #1}, #2}
\def \MEM #1 #2 {{\em Mem. R. Astr. Soc.\/} {\bf #1}, #2}
\def \PLR #1 #2 {{\em Phys. Lett. Rev.\/} {\bf #1}, #2}
\def \PASJ #1 #2 {{\em Publ. Astron. Soc. Japan\/} {\bf #1}, #2}
\def \PASP #1 #2 {{\em Publ. Astr. Soc. Pacific\/} {\bf #1}, #2}
\def \NAT #1 #2 {{\em Nature\/} {\bf #1}, #2}

\documentstyle{memsait}
\input epsf.sty
\begin{opening}
\title{EMISSION PROCESSES IN GAMMA--RAY BURSTS}
\author{GABRIELE GHISELLINI}
\institute{Osservatorio Astronomico di Brera, Merate, Italy}
\date{}             
\end{opening}

\begin{document}

\oddpagefooter{}{}{} 
\evenpagefooter{}{}{} 
\ 
\bigskip

\begin{abstract}
Recent results of the hectic research activity about gamma--ray bursts will
be reviewed, with emphasis about the emission processes at the origin of
the observed $\gamma$--rays.
The conventional synchrotron shock scenario is found to have problems,
due to the very short cooling times of the emitting electrons, which implies
a predicted spectrum, $F_{\nu}\propto \nu^{-1/2}$, much steeper than what 
is observed.
It is therefore compelling to look for alternative emission processes, 
such as quasi--thermal Comptonization, implying the presence of 
mildly or sub--relativistic electrons, producing, through multiple 
Compton scatterings, a spectrum $\propto \nu^0$ ending with a Wien 
peak where photons and electron energies balance.
The afterglow light, instead, can be indeed due to synchrotron radiation, 
and a confirmation of this is the recently detected optical linear
polarization of the afterglow of GRB 990510.
Some consequences of this discovery will be outlined.
A quantum leap in our understanding of the physics of gamma--ray bursts
is expected to come with SWIFT, a NASA--MIDEX dedicated satellite.
\end{abstract}

\section{Introduction}

After 30 years since their detection by the VELA satellites,
we just now start to understand the physics of gamma--ray bursts (GRB).
This has been made possible by the precise location of the
Wide Field Camera (WFC) of $Beppo$SAX, which allowed the detection of their
X--ray afterglow emission (Costa et al. 1997) and the optical follow up
observations, leading to the discovery that they are cosmological sources 
(van Paradijs et al. 1997).
The huge energy and power releases required by their cosmological
distances support the fireball scenario (Cavallo \& Rees 1978; Rees
\& M\'esz\'aros 1992; M\'esz\'aros \& Rees 1993),
even if we do not know yet which kind of progenitor makes the GRB phenomenon.

The most accepted picture for the burst and afterglow emission is the
internal/external shock scenario (Rees \& M\'esz\'aros 1992;
Rees \& M\'esz\'aros 1994; Sari \& Piran 1997).
According to this scenario, the burst emission is due to
collisions of pairs of relativistic shells (internal shocks), while
the afterglow is generated by the collisionless shocks produced by shells
interacting with the interstellar medium (external shocks).

The $\gamma$--ray light curves show an extremely variable emission,
with variability timescales as short as a few milliseconds.
This, coupled with the huge powers involved, requires that
the emitting plasma is moving at relativistic speeds, with bulk Lorentz
factors $\Gamma>100$, to avoid strong suppression of high
energy $\gamma$--rays due to photon--photon collisions.
Indeed, all the radiation we see is believed to come from the 
transformation of ordered kinetic energy of the fireball into random energy.
This must however happen at some distance from the explosion site,
to allow the shells to be transparent to the produced radiation.
Fiducial numbers for these distances are $R\simeq 10^{12}$--$10^{13}$ cm
for the internal shocks, and $R\simeq 10^{16}$ cm for external ones.

It is reasonable to assume that internal and external shocks can amplify
seed magnetic fields and accelerate electrons to relativistic random energies.
If this is the case, then the main radiation mechanism is the synchrotron
process, responsible for both the $\gamma$--ray and for the afterglow emission.
There are strong evidences that this is the main process
operating during the afterglow: the power law decay of the flux
with time, the observed power law energy spectra (for reviews see 
Piran 1999; M\'esz\'aros 1999), and the recently detected linear optical 
polarization in GRB 990510 (Covino et al. 1999, Wijers et al. 1999).

For the bursts itself, the main evidence is the prediction of the
typical energy at which the observed spectrum peaks, in a
$\nu F_\nu$ representation.
However, the very same synchrotron shock scenario (ISS) inevitably 
predicts very fast radiative cooling of the emitting particles,
with a resulting severe disagreement between the predicted and the observed
spectrum (Ghisellini, Celotti \& Lazzati 1999, see also Cohen et al. 1997;
Sari, Piran \& Narayan 1998; Chiang 1999).

In the following, after having briefly recalled some basic facts concerning 
GRBs, I will discuss the problems with the synchrotron interpretation of the 
$\gamma$--ray emission and some possible alternatives.

Another hot issue in the GRB field is the possible collimation of their
emitting plasma, leading to anisotropic emission
able to relax the power requirements, at the expense of an increased
burst event rate.
I will show that in this respect polarization studies could be crucial, 
since there can be a link between the deceleration of a {\it collimated} 
fireball, the time behavior of the polarized flux and its
position angle, and the light curve of the total flux.

\section{Some Facts}

\noindent
{\it Duration ---} 
The duration distribution is clearly bimodal.
{\it Short} burst durations range between 0.01 to 2 seconds.
{\it Long} burst last from 2 to a few hundred seconds.
All information derived from the precise location of GRBs refer to 
{\it long} bursts.
There is some indication that short bursts may have a different
$\log N - \log S$ (Tavani 1998).

\vskip 0.2 true cm
\noindent
{\it Spectra and fluences ---} 
Most $\gamma$--ray fluences (i.e. the flux integrated over the duration
of the burst) are in the range $10^{-6}$--$10^{-4}$ erg cm$^{-2}$.
GRB spectra are very hard, and can be fitted by two smoothly connected
power laws, implying a peak (in a $\nu$--$\nu F_\nu$ plot) at 
an energy $\nu_{\rm peak}$ of a few hundreds keV.
Below $\nu_{\rm peak}$ the spectrum has a slope $\alpha$ in the range
[$-$1, $+$0.5] (Preece et al. 1998, Lloyd \& Petrosian 1999; 
$F_\nu \propto \nu^{-\alpha}$).
This spectral index varies during the burst, as does $\nu_{\rm peak}$.
The general trend is that the spectrum softens, and $\nu_{\rm peak}$
decreases, with time. 
More precise statements must however wait for larger area detectors,
since what we inevitably do, at present, is to fit a time integrated 
spectrum of a very rapidly variable source:
the minimum integration time is $\sim$1 second for the strongest bursts,
while the variability timescales can be hundreds of times shorter.
Some bursts have been detected at very large $\gamma$--rays energies 
($>$ 100 MeV) (see the review by Fishman \& Meegan 1995, and references
therein).

\vskip 0.2 true cm
\noindent 
{\it X--ray afterglow ---} 
Nearly all of the burst detected by the BeppoSAX GRBM and WFC,
and that could be followed with the Narrow Field Instruments
of BeppoSAX a few (6--10) hours after the event, showed an X--ray afterglow,
whose flux decays in time as a power law.

\vskip 0.2 true cm
\noindent 
{\it Optical afterglow ---} 
For about 2/3 of the bursts with good locations an optical afterglow
has been detected. 
The monochromatic flux decreases in time as a power law 
$F_\nu (t)\propto t^{-\delta}$, with $\delta$ in the range 0.8--2.
Remarkably, GRB 990123 has been detected by the robotic telescope
ROTSE 22 seconds after the $\gamma$--ray 
trigger at $m\sim 11.7$, reaching $m\sim 8.97$ 47 seconds after the trigger
(Akerlof \& McKay 1999).
Usually, the magnitudes of the optical afterglow detected
a few hours from the event are in the range 18--21.

\vskip 0.2 true cm
\noindent 
{\it Radio afterglow --- }
A few bursts showed radio emission, usually some time ($\sim$days) after
the burst event.
Violent radio flux variations in GRB 970508 (Frail et al. 1997) and 
GRB 980329 (Taylor et al. 1998) have been interpreted as due to interstellar 
scintillation, effective if the source is extremely compact.
Transition from the phase of violent activity to a more quiescent phase
has been interpreted as due to the expansion of the radio emitting source,
beautifully confirming the relativistic expansion hypothesis.

\vskip 0.2 true cm
\noindent 
{\it Redshifts --- }
Up to July 15, 1999, we know the redshift of 8 GRBs, with an additional
one (GRB 980329) estimated by a cut-off in the spectrum interpreted as 
$L_\alpha$ absorption (Fruchter 1999a), and another one (GRB 980425) 
of controversial identification with the supernova SN1998bw.
In Table 1 we list some observed characteristics of the GRBs of known
redshifts, and Fig. 1 shows the power emitted in $\gamma$--rays, the
fluences and the magnitude of the host galaxy candidates as a function of
redshift.

\begin{figure}
\vskip -0.5 true cm
\epsfysize=9cm 
\hspace{1.5cm}\epsfbox{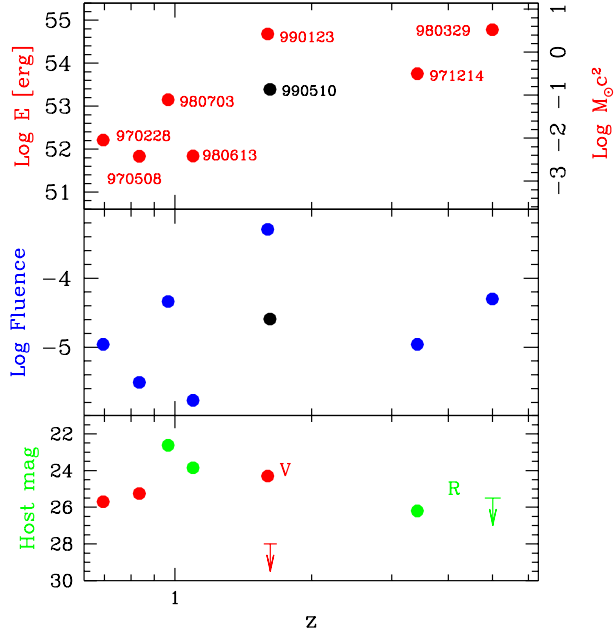} 
\vskip -0.5 true cm
\caption[h]{{\it Top panel:} Total energy (in $\gamma$--rays only) calculated
under the assumption of isotropic emission for the bursts of known redshift,
as a function of redshift. We omitted GRB 980425 because 
the possible identification with the ``nearby" ($\sim$40 Mpc) 
SN1998bw is controversial.
Note that the most powerful bursts, if isotropic, require the complete
conversion into energy of more than a solar mass (right scale).
{\it Medium panel:} $\gamma$--ray fluences as a function of redshift.
No trend can be seen.
{\it Bottom panel:} Magnitudes of the host galaxies as a function of redshift.
A great diversity seems to be present: note the upper limit ($R>28$) for 
the host galaxy of GRB 990510 (Fruchter et al. 1999d).}
\end{figure}

\vskip 0.3 true cm
\centerline{\bf Tab. 1 - Gamma--ray burst of known redshift}
\begin{table}[h]
\vskip -0.4 true cm
\hspace{1.5cm} 
\begin{tabular}{|l|c|c|c|c|c|c|}
\hline
\hline
GRB            &z     &Ref. &Fluence$^a$ &$m_{\rm host}$ &Ref. &WFC$^b$ \\
\hline
970228         &0.695  & 1 &1.1 (GRBM) &$R=25.3$   &12     &4.2   \\
970508         &0.835  & 2 &0.3    &$R=25.1$       &13     &1.1   \\
971214         &3.418  & 3 &1.1    &$R=26.2$       &14     &1.0   \\
980329         &5?     & 4 &5.0    &$R>25.5$       &15     &6.0   \\
980425         &0.0083 & 5 &0.4    &$R=14.3$       &16     &3.0   \\
980613         &1.0964 & 6 &0.2    &$R=24.1$       &17     &0.6   \\
980703         &0.966  & 7 &4.6    &$R=22.6$       &18     &---   \\
990123         &1.600  & 8,9 &51   &$R=23.7$       &19     &3.4   \\
990510         &1.619  & 10 &2.6   &$V>28$         &20     &4.3   \\
990712         &0.430  & 11 &---   &---            &---    &---   \\
\hline
\hline
\end{tabular}
\end{table}
\vskip -0.3 true cm
\noindent
{\it Notes:} $a$: BATSE $\gamma$--ray fluences in units of $10^{-5}$ 
erg cm$^{-2}$. $b$: WFC flux in Crab units.
{\it References}: 
1: Djorgovski et al. 1999b;
2: Metzger et al. 1997;
3: Kulkarni et al. 1998b
4: Fruchter 1999a;                
5: Tinney et al. 1998;
6: Djorgovski et al. 1999a;
7: Djorgovski et al. 1998a;
8: Kelson et al. 1999;
9: Hjorth et al. 1999;
10: Vreeswijk et al. 1999;
11: Galama et al. 1999;
12: Fruchter et al. 1999b;
13: Bloom et al. 1998a;
14: Odewann et al. 1998;
15: Djorgovski et al. 1998a;
16: Kulkarni et al. 1998b;
17: Djorgovski et al. 1998c;
18: Bloom et al. 1999b;
19: Fruchter et al. 1999c;
20: Fruchter et al. 1999d.

\section{The internal shock scenario}

The energy involved in $\gamma$--ray burst explosions is huge.
No matter in which form the energy is initially injected,
a quasi--thermal equilibrium between matter and radiation is reached,
with the formation of electron--positron pairs accelerated to
relativistic speeds by the high internal pressure.
This is a {\it fireball}.
When the temperature of the radiation (as measured in the comoving
frame) drops below $\sim$50 keV the pairs annihilate 
faster than the rate at which are produced.
But the presence of even a small amount of barions, corresponding to 
only $\sim 10^{-6}~ M_\odot$,
makes the fireball opaque to Thomson scattering: the internal radiation
thus continues to accelerate the fireball until most of its initial 
energy has been converted into bulk motion. 
After this phase the fireball expands at a constant speed and at some point 
becomes transparent.
If the central engine is not completely impulsive, but works intermittently,
it can produce many shells (i.e. many fireballs)
with slightly different Lorentz factors.
Late but faster shells can catch up early slower ones,
producing shocks which give rise to the observed burst emission.
In the meantime, all shells interact with the interstellar medium, and 
at some point the amount of swept up matter is large enough to decelerate
the fireball and produce other radiation which can be identified with 
the afterglow emission observed at all frequencies.

\section{Typical synchrotron frequency}

In the comoving frame of the faster shell, protons of the other shell
have an energy density 
$U^\prime_{\rm p}=(\Gamma^\prime-1)n^\prime_{\rm p} m_{\rm p}c^2$,
where $\Gamma^\prime\sim 2$ is the bulk Lorentz factor of the slower shell
measured in the rest frame of the other. 
The magnetic energy density $U^\prime_{\rm B}$ can be amplified to values 
close to equipartition with the proton energy density, 
$U^\prime_{\rm B} = \epsilon_{\rm B} U^\prime_{\rm p}$.
The proton density of the shell can be estimated by its kinetic power:
$L_{\rm s} = 4\pi R^2 \Gamma^2 c n^\prime_{\rm p}  m_{\rm p}c^2$,
yielding $B = (2\epsilon_{\rm B} L_{\rm s}/ c)^{1/2}/( R\Gamma)$.
Also each electron can share a fraction of the available energy, and
if there is one electron for each proton, namely {\it if electron--positron
pairs are not important}, then
$\gamma m_{\rm e} c^2 = \epsilon_{\rm e} m_{\rm p} c^2$.
These simple hypotheses lead to a predicted observed synchrotron
frequency 
\begin{equation}
h\nu_{\rm s}\, \sim\, 4\, \epsilon_{\rm e}^2 \epsilon_{\rm B}^{1/2} 
{L_{\rm s, 52}^{1/2} \over R_{13} (1+z) }\,\, {\rm MeV}
\end{equation}
in very good agreement with observations.
Note that the `equipartition coefficients', $\epsilon_{\rm B}$ and 
$\epsilon_{\rm e}$, must be close to unity for the
observed value of $\nu_{\rm peak}$ to be recovered.
In turn this also implies/requires that pairs
cannot significantly contribute to the lepton density.

\section{Cooling is fast}

The synchrotron process is a very efficient radiation process.
With the strong magnetic fields required to produce the observed
$\gamma$--rays, the synchrotron cooling time is therefore very short.
As pointed out by Ghisellini, Celotti \& Lazzati (1999), the cooling
time (in the observer frame) can be written as:
\begin{equation}
t_{\rm cool}\,\sim \, 10^{-7} {\epsilon_{\rm e}^3 \Gamma_2 \over 
\nu_{\rm MeV}^2 (1+U_{\rm r}/U_{\rm B})(1+z)}\, \, {\rm s},
\end{equation}
where $U_{\rm r}$ is the radiation energy density.
Since the shortest integration times are of the order of 1 s, the observed 
spectrum is {\it always} the time integrated spectrum produced by
a rapidly cooling particle distribution.

Since $t_{\rm cool}\propto 1/\gamma$, in order to conserve the particle
number, the instantaneous cooling distribution has to satisfy 
$N(\gamma, t) \propto 1/\gamma$.
When integrated over time, the contribution from particles with
different Lorentz factors is `weighted' by their cooling timescale 
$\propto 1/\gamma$.
Therefore the predicted (integrated) flux spectrum is 
$F_\nu \propto t_{\rm cool} N(\gamma)\dot\gamma d\gamma/ d \nu
\propto \nu^{-1/2}$,
extending from $\sim$1 keV to $h\nu_{\rm peak}\sim$MeV energies.
We thus conclude that, within the assumptions of the ISS, a major
problem arises in interpreting the observed spectra as synchrotron
radiation.
Ghisellini, Celotti \& Lazzati (1999) have discussed possible `escape
routes', such as deviations from equipartition, fastly changing magnetic
fields, strong cooling by adiabatic expansion and re--acceleration of
the emitting electrons.
None of these possibilities help. 
The drawn conclusion is that the burst emission is probably produced by 
another radiation process.
An alternative is quasi thermal Comptonization (Ghisellini \& Celotti
1999; Liang 1997, Thompson 1994).

\section{The typical peak frequency of GRBs}

Why do GRB spectra peak at a few hundreds keV?
This must be due to a quite robust mechanism and/or a feedback process.
The fact that $h\nu_{\rm peak}$ is close to $m_{\rm e}c^2$
is tempting. 
I discuss below three possibilities.

\subsection{``Pair feedback"}

Assume that the main emission process is quasi thermal Comptonization
(Ghisellini \& Celotti 1999) by a distribution of electrons (and positrons) 
peaked at subrelativistic energies (not necessarily thermal).
Photons produced above the pair production threshold get absorbed and
create pairs.
This process has been extensively studied in the past
assuming thermal particles, steady state and pair equilibrium 
(equal pair production and annihilation rates).
The main result is that for compact sources
the maximum temperature is constrained in a narrow range
[30--300 keV ](Svensson 1982, 1984): for increasing luminosities 
the equilibrium pair density increases, more particles are then sharing the 
available energy, making the temperature to decrease.
If an high energy tail is present, more photons are created above 
the threshold for pair production with respect to the case
of a pure Maxwellian, and thus pairs become important for 
temperatures smaller than in the completely thermal case 
(see e.g. Stern 1999).

\subsection{``Brainerd break"}

Brainerd (1994) linked the typical high energy cut--off of GRBs to
the effect of down--scattering: photons with energies much 
larger than $m_{\rm e}c^2$ pass undisturbed through a scattering medium 
because of the reduction with energy of the Klein Nishina
cross section, while photons with energies just below $m_{\rm e}c^2$ 
interact, and their energy after the scattering is reduced.
The net effect is to produce a ``downscattering hole" in the spectrum,
between $\sim m_{\rm e}c^2/\tau^2$ and $\sim \tau m_{\rm e}c^2$.
The attractive feature of this model is that the cut--off energy is
associated with $m_{\rm e}c^2$.
The difficulty is that a significant part of the power originally 
radiated by the burst goes into heating (by the Compton process) 
of the scattering electrons.

\subsection{Pair production break}

Suppose that the circumburst material has a density $n_{\rm ext}$
and that, close to the burst emission site (i.e. between $R$ and $2R$),
the scattering optical depth is $\tau_{\rm ext}$.
This material scatters back a fraction $\tau_{\rm ext}L$ of the 
burst power. 
If we require that the primary spectrum is not modified by photon--photon 
absorption, the optical depth of the scattering matter and its density must be
(Ghisellini \& Celotti 1999)
\begin{equation}
\tau_{\rm ext} \, < \, 3.7\times 10^{-9} {R_{13}\over L_{50}} 
\, \to \, n_{\rm ext} \, <\, {5.5\times 10^{2} \over L_{50}}\, \, 
{\rm cm^{-3}}
\end{equation}
This requirement is particularly severe, especially in the case of bursts 
originating in dense stellar forming regions.
On the other hand, photon--photon opacity may be an important
ingredient to shape the spectrum, and {\it the} reason why 
GRB spectra peak at around 300 keV.
The pairs created ahead of the fireball radiate their energy in a short time, 
and are re--accelerated by the incoming fireball.
The net effect may be simply to increase the density of the
radiating particles, introducing a feedback process: an increased
density lowers the effective temperature $\to$ less energy is radiated
above $m_ec^2$ $\to$ the number of pairs produced via the ``mirror" 
process decreases $\to$ the new pair density decreases, and so on.
Another feedback is introduced by the fact that the photons
scattered back to the emitting shell will increase the number of
seed photons and the radiative cooling rate, softening the spectrum.

\section{Polarized afterglows}

Covino et al. (1999) and Wijers et al. (1999) detected a small, but highly
significant, degree of linear polarization of the optical afterglow
of GRB 990510: $P=(1.7\pm0.2)$\% 18.5 and 20.5 hours after the burst,
respectively.
\begin{figure}
\epsfysize=5cm 
\hspace{2.5cm}\epsfbox{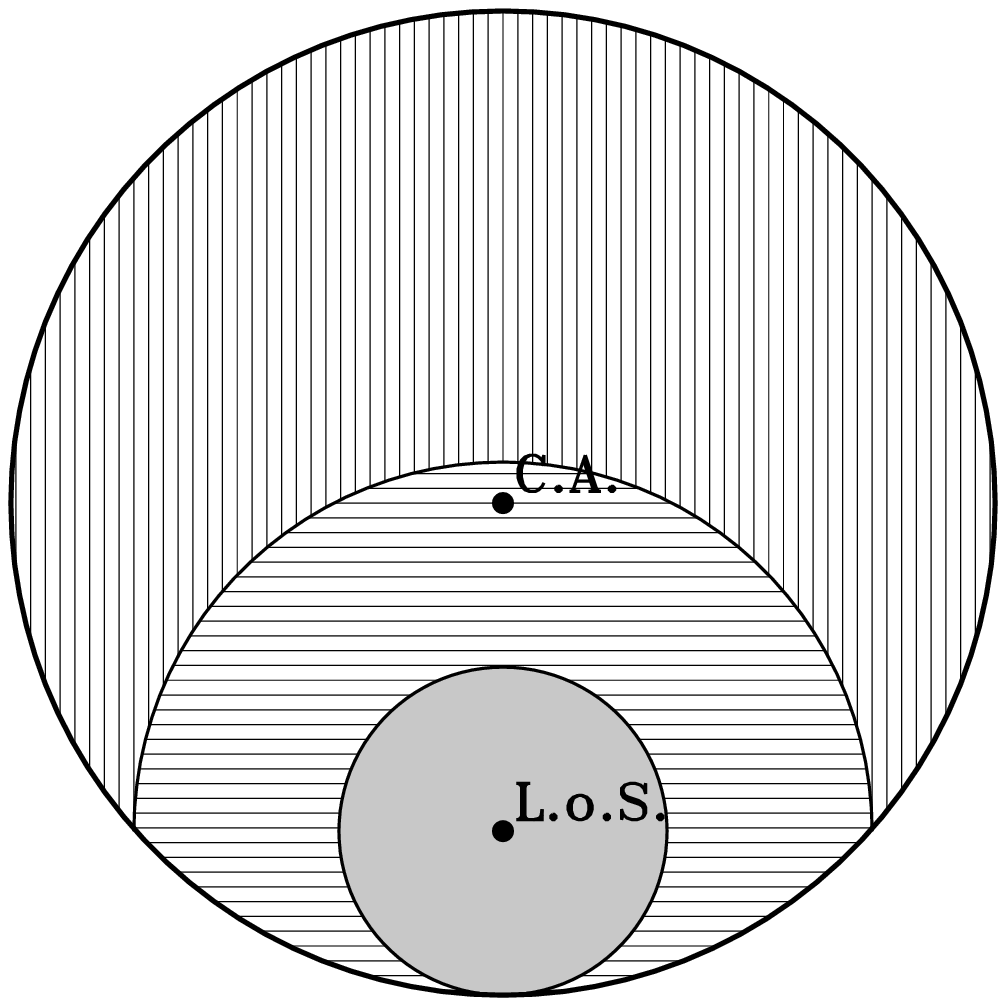} 
\caption[h]{Front view of the beamed fireball:
C.A. stands for Cone Axis, while L.o.S. stands for
Line of Sight.
The grey shaded area produces unpolarized radiation, due to 
the complete symmetry around the line of sight. 
The horizontal line shaded region 
produces a horizontal component of polarization while the upper region 
(vertically line shaded) produces vertical polarization.
From Ghisellini \& Lazzati, 1999).} 
\end{figure}

\begin{figure}
\epsfysize=8.5cm 
\hspace{2.5cm}\epsfbox{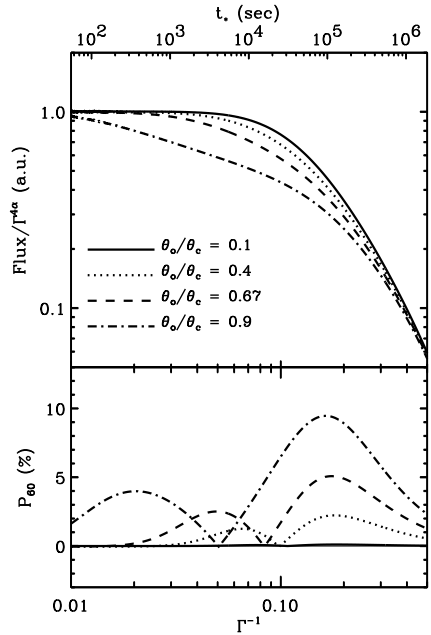} 
\caption[h]{Lightcurves of the total flux (upper panel) and of the
polarized fraction (bottom panel) for four different choices
of the ratio $\theta_o/\theta_c$. 
The cone aperture angle $\theta_c=5^\circ$. 
$\theta_o$ is the viewing angle.
The lightcurve of the total flux assumes a constant spectral index
$\alpha=0.6$ for the emitted radiation. 
Note that the highest polarization values are associated with
total flux lightcurves steepening more gently.
From Ghisellini \& Lazzati, 1999.} 
\end{figure}

To produce some polarized light some {\it asymmetry} is required.
In particular, if the radiation is due to the synchrotron process,
the magnetic field cannot be completely tangled, but must have some
degree of order.

Prior of the observational discovery, there have been theoretical studies 
predicting a larger degree of polarization 
($P\sim 10\%$: Gruzinov \& Waxman 1999)
due to causally disconnected region of maximally ordered magnetic field.
Since the number $N$ of these regions is limited, the resulting polarization
is of the order of $60\%/\sqrt{N}$, where 60\% is the value reached in each
single subregion.
This model requires a perfectly ordered magnetic field,
on a scale that increases in time at almost the speed of light.

We (Ghisellini \& Lazzati 1999) have instead considered an alternative 
scenario, in which the required asymmetry is due to field compression and
light aberration.
Following Laing (1980), assume to compress a region embedded in an
completely tangled magnetic field. 
After compression, the region becomes a slab: the field
is ``squeezed" in one direction, and appears again completely
random for face--on observes (with respect to the direction of compression),
but highly ordered to edge--on observers.
Photons emitted in the plane of the slab can then be highly polarized.
If the slab moves with Lorentz factor $\Gamma$,
those photons emitted in the slab plane (perpendicularly to the
direction of motion) are aberrated in the observer frame, and make an angle
$\theta =1/\Gamma$ with respect to the slab velocity.
Observer looking at the moving slab at this angle will detect a large degree of
optical polarization.

Consider a fireball collimated into a cone of semi--aperture
angle $\theta_{\rm c}$.
Let $\theta_{\rm o}$ be the angle between the cone axis and the line of sight.
As long as $1/\Gamma < \theta_{\rm c}/\theta_{\rm o}$, the observer
receive radiation from a circle entirely contained in the jet.
There is no asymmetry, and no polarization.
But when $1/\Gamma > \theta_{\rm c}/\theta_{\rm o}$ (see Fig. 2),
some position angles are not canceled, and some polarization survives.
The degree of polarization as a function of time is shown in Fig. 3:
Note the two maxima (a third could be present if one considers 
side expansion of the jet which we neglected, see Sari 1999).
The position angles of the two maxima are orthogonal between each other.
Fig. 3 also shows the light curve of the total flux, to illustrate the tight
link that this model predicts between the existence of polarization and
the gradual steepening of the light curve.

Should these ideas be confirmed, we would have a very powerful tool to 
know the degree of collimation of the fireball, and hence the true total 
emitted power.

\section{SWIFT: a dedicated satellite}
Our knowledge of gamma--ray bursts had a quantum leap in the last two years.
But we do not know yet what is their main cause (i.e. their progenitors,
see the review by Vietri 1999, this volume),
if it is the same for short and long bursts, what is the main radiation
process of the burst, why their spectrum peaks at a well defined energy.
But the information in hand lead us to expect that they could be perfect
cosmological probes, doing, for redshifts greater than 5, what quasars
have done for lower redshifts.
Therefore there is a double interest to pursue GRB research: to understand
their nature and to use them to test the distant universe.

For these reasons another quantum leap is necessary, and it is foreseen to come
from the launch of a dedicated NASA--MIDEX 
satellite, called SWIFT, able to detect
more than 100 burst per year and to slew to them with X--ray and optical--UV
telescopes {\it in less than 50 seconds!}
(see the dedicated web page at {\it http://swift.gsfc.nasa.gov/}).
With these capabilities, we will be able to have a large sample of 
accurate locations, and then, after the optical follow--up
from the ground, we will have a large number of redshifts.
We will study the relatively soft X--ray emission [0.1--10 keV]
and the optical emission while the burst is still on, and we will know 
the transition between the burst proper and the afterglow.
Definitive approval of the mission is expected in the fall of 1999,
and launch is foreseen for the year 2003.
The Italian community is deeply involved and will provide the
X--ray mirrors, developed by the Brera Observatory, and the
Malindi ground station, developed by ASI for {\it Beppo}SAX.
We think that this mission, besides being a cornerstone for the study
of gamma--ray bursts, will be a great opportunity for the Italian 
community at large.

\acknowledgements
I thank Annalisa Celotti and Davide Lazzati for a very productive collaboration.


\end{document}